# Functional Observers with Linear Error Dynamics for Nonlinear Systems

Costas Kravaris* and Sunjeev Venkateswaran

*Texas A&M University, College Station, TX 77843-3122 USA*

*Abstract:* In this work, the problem of designing observers for estimating a single nonlinear functional of the state is formulated for general nonlinear systems. Notions of functional observer linearization are also formulated, in terms achieving exactly linear error dynamics in transformed coordinates and with prescribed rate of decay of the error. Necessary and sufficient conditions for the existence of a lower-order functional observer with linear dynamics are derived. The results provide a direct generalization of Luenberger's linear theory of functional observers to nonlinear systems.

1. INTRODUCTION

The problem of estimating a function of the state vector, without the need of estimating the entire state vector, arises in many applications. The design of output feedback controllers based on a state feedback design is a classical example, where it is only the state feedback function that needs to be estimated and not the entire state vector. Another class of applications is related to the design of inferential control systems, where one output is measured and a different output, which is unmeasured, needs to be regulated. From a practical point of view, the most common class of applications is related to condition monitoring of dynamic systems.

These applications motivate the development of functional observers, the aim being a reduction of dimensionality relative to a full-state observer. The notion of a functional observer was first defined in Luenberger's pioneer work on observers for linear multivariable systems[1, 2]. Luenberger proved that for a linear system, it is feasible to construct a functional observer with number of states equal to observability index minus one.

The basic theory of linear functional observers can be found in standard linear systems texts, e.g.[3]. In recent years, there has been a renewed interest in linear functional observers [4-9], the goal being to find the smallest possible order of the linear functional observer.

For nonlinear systems, there have been significant developments in the theory of full-state observers, with a variety of methods and approaches. In particular, in the context of exact linearization methods [10-17], Luenberger theory for full-state observers has been extended to nonlinear systems in a direct and analogous manner. The main goal of the present work is to develop a direct generalization of Luenberger's functional observers to nonlinear systems.

---

*author to whom correspondence should be addressed. Email: kravaris@tamu.edu

The present work studies the design of functional observers for nonlinear systems from the point of view of exact observer error linearization. Necessary and sufficient conditions for the existence of functional observers with linear error dynamics are derived, leading to simple formulas for observer design with eigenvalue assignment. In particular, in the present work, we consider unforced nonlinear systems of the form

$$\frac{dx}{dt} = F(x)$$
$$y = H(x) \qquad (1.1)$$
$$z = q(x)$$

where:

$x \in \mathbb{R}^n$ is the system state
$y \in \mathbb{R}^p$ is the vector of measured outputs
$z \in \mathbb{R}$ is the (scalar) output to be estimated

and $F: \mathbb{R}^n \to \mathbb{R}^n$, $H: \mathbb{R}^n \to \mathbb{R}^p$, $q: \mathbb{R}^n \to \mathbb{R}$ are smooth nonlinear functions. The objective is to construct a functional observer of order $v < n$, which generates an estimate of the output z, driven by the output measurement y.

Section 2 will define the notion of functional observer for a system of the form (1.1) in a completely analogous manner to Luenberger's definition for linear systems. Section 3 will pose the problem of functional observer design and point out its challenges. Section 4 will define notions of exact linearization for the functional observer problem, in the same vein as they have been defined for full-state observers in the literature. Section 5 will develop necessary and sufficient conditions for the solution of the functional observer linearization problem, when both the observer dynamics and its output must be linear, as well as a simple formula for the resulting functional observer. The results of Section 5 will be specialized to linear time-invariant systems in Section 6, leading to simple and easy-to-check conditions for the design of lower-order functional observers. In Section 7, necessary and sufficient conditions for solvability of a more general functional observer linearization problem will be derived.

2. DEFINITION OF A FUNCTIONAL OBSERVER FOR A NONLINEAR DYNAMIC SYSTEM

In complete analogy to Luenberger's construction for the linear case, we seek for a mapping

$$\xi = \mathcal{T}(x) = \begin{bmatrix} \mathcal{T}_1(x) \\ \vdots \\ \mathcal{T}_v(x) \end{bmatrix} \qquad (2.1)$$

from $\mathbb{R}^n$ to $\mathbb{R}^v$, to immerse system (1.1) to a $v$-th order system ($v < n$), with input y and output z:

$$\frac{d\xi}{dt} = \varphi(\xi, y)$$
$$z = \omega(\xi, y) \qquad (2.2)$$

where $\varphi: \mathbb{R}^n \times \mathbb{R}^p \to \mathbb{R}^\nu$, $\omega: \mathbb{R}^n \times \mathbb{R}^p \to \mathbb{R}$, the aim being that system (2.2), driven by the measured output y of (1.1), can generate an estimate of unmeasured output z of (1.1).

But in order for system (1.1) to be mapped to system (2.2) under the mapping $\mathcal{T}(x)$, the following relations have to hold:

$$\frac{\partial \mathcal{T}}{\partial x}(x)F(x) = \varphi\big(\mathcal{T}(x), H(x)\big) \tag{2.3}$$

$$q(x) = \omega\big(\mathcal{T}(x), H(x)\big) \tag{2.4}$$

The foregoing considerations lead to the following definition of a functional observer:

<u>Definition 1</u>: *Given a dynamic system (1.1), with* y *being the vector of measured outputs and* z *is the scalar output to be estimated, the system*

$$\begin{aligned} \frac{d\hat{\xi}}{dt} &= \varphi(\hat{\xi}, y) \\ \hat{z} &= \omega(\hat{\xi}, y) \end{aligned} \tag{2.5}$$

*where* $\varphi: \mathbb{R}^n \times \mathbb{R}^p \to \mathbb{R}^\nu$, $\omega: \mathbb{R}^n \times \mathbb{R}^p \to \mathbb{R}$ $(\nu < n)$ *is called a functional observer for (1.1), if in the series connection*

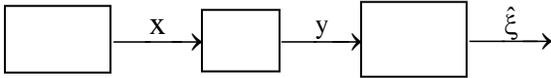

*the overall dynamics*

$$\begin{aligned} \frac{dx}{dt} &= F(x) \\ \frac{d\hat{\xi}}{dt} &= \varphi(\hat{\xi}, H(x)) \end{aligned} \tag{2.6}$$

*possesses an invariant manifold* $\hat{\xi} = \mathcal{T}(x)$ *with the property that* $q(x) = \omega\big(\mathcal{T}(x), H(x)\big)$.

In the above definition, the requirement that $\hat{\xi} = \mathcal{T}(x)$ is an invariant manifold of (2.6), i.e. that

$\hat{\xi}(0) = \mathcal{T}(x(0)) \implies \hat{\xi}(t) = \mathcal{T}(x(t)) \;\; \forall t > 0$, translates to $\dfrac{\partial \mathcal{T}}{\partial x}(x)F(x) = \varphi\big(\mathcal{T}(x), H(x)\big)$, which is condition (2.3) stated earlier.

If the functional observer (2.5) is initialized consistently with the system (1.1) i.e. if
$\hat{\xi}(0) = \mathcal{T}(x(0))$, then $\hat{\xi}(t) = \mathcal{T}(x(t))$, and therefore
$\hat{z}(t) = \omega\big(\hat{\xi}(t), y(t)\big) = \omega\big(\mathcal{T}(x(t)), H(x(t))\big) = q(x(t)) \;\; \forall t > 0$

which means that the functional observer will be able to exactly reproduce z(t).

In the presence of initialization errors, additional stability requirements will need to be imposed on the $\hat{\xi}$-dynamics, for the estimate $\hat{z}(t)$ to asymptotically converge to z(t).

At this point, it is important to examine the special case of a linear system, where $F(x) = Fx$, $H(x) = Hx$, $q(x) = qx$ with F, H, q being $n \times n$, $p \times n$, $1 \times n$ matrices respectively, and a linear mapping $\mathcal{T}(x) = Tx$ is considered. Definition 1 tells us that for a linear time-invariant system

$$\frac{dx}{dt} = Fx$$
$$y = Hx \qquad (2.7)$$
$$z = qx$$

the system

$$\frac{d\hat{\xi}}{dt} = A\hat{\xi} + By \qquad (2.8)$$
$$\hat{z} = C\hat{\xi} + Dy$$

will be a functional observer if the following conditions are met:

$$TF = AT + BH \qquad (2.9)$$
$$q = CT + DH \qquad (2.10)$$

for some $\nu \times n$ matrix T. These are exactly Luenberger's conditions for a functional observer for a linear time-invariant system [2].

3. DESIGNING LOWER-ORDER FUNCTIONAL OBSERVERS FOR NONLINEAR SYSTEMS

In order to design of a functional observer for a nonlinear system, one must be able to find a mapping $\mathcal{T}(x) = \begin{bmatrix} \mathcal{T}_1(x) \\ \vdots \\ \mathcal{T}_\nu(x) \end{bmatrix}$ to satisfy conditions (2.3) and (2.4), i.e. such that:

- $\frac{\partial \mathcal{T}_j}{\partial x}(x)F(x)$, $j = 1, \cdots, \nu$ is a function of $\mathcal{T}_1(x), \cdots, \mathcal{T}_\nu(x)$, $H(x)$
- $q(x)$ is a function of $\mathcal{T}_1(x), \cdots, \mathcal{T}_\nu(x)$, $H(x)$

However, such scalar functions $\mathcal{T}_1(x), \cdots, \mathcal{T}_\nu(x)$ may not exist if $\nu < n - p$.

Moreover, even when they do exist, there is an additional very important requirement:

Since $\frac{\partial \mathcal{T}}{\partial x}(x)F(x) = \varphi(\mathcal{T}(x), H(x))$ will define the right-hand side of the functional observer's dynamics, it must be such that the functional observer's dynamics is stable and the decay of the error is sufficiently rapid.

This paper will address the functional observer design problem, focusing on *finding conditions under which low-order functional observers are feasible*.

## 4. EXACT LINEARIZATION OF A FUNCTIONAL OBSERVER

The concept of exact observer linearization has been formulated in the literature for full-state observers. We will start this section with a brief necessary review, following [10], [11], [12]. Subsequently we will propose an extension of the concept of exact linearization to functional observers, and we will discuss both the inherent similarity and the fundamental difference with exact linearization of full-state observers.

Consider a nonlinear system

$$\frac{dx}{dt} = F(x) \qquad (4.1)$$
$$y = H(x)$$

If a mapping $\xi = \mathcal{T}(x)$ from $\mathbb{R}^n$ to $\mathbb{R}^\nu$ can be found to map system (4.1) to a linear system

$$\frac{d\xi}{dt} = A\xi + By \qquad (4.2)$$

where A and B are $\nu \times \nu$ and $\nu \times p$ matrices respectively, the idea is to use system (4.2) as the basis for a state observer. The mapping $\mathcal{T}(x)$ must satisfy

$$\frac{\partial \mathcal{T}}{\partial x}(x) F(x) = A \mathcal{T}(x) + B H(x) \qquad (4.3)$$

Assuming for the moment that the partial differential equation (4.3) can be solved, *it will be possible to reconstruct the state if the mapping* $x \mapsto \xi$ *is injective or if* $x \mapsto (\xi, y)$ *is injective.* The observer will then consist of a replica of (4.2)

$$\frac{d\hat{\xi}}{dt} = A\hat{\xi} + By \qquad (4.4)$$

along with an algebraic equation to calculate the state estimate. In particular,

-- for $\nu = n$ (full-order observer*)*, the state estimate will be calculated as $\hat{x} = \mathcal{T}^{-1}(\hat{\xi})$,

-- for $\nu = n - p$ (reduced-order observer*)*, the state estimate will be the solution of $\begin{Bmatrix} \mathcal{T}(\hat{x}) = \hat{\xi} \\ H(\hat{x}) = y \end{Bmatrix}$.

In both cases, the observer's error dynamics will follow

$$\frac{d}{dt}\left(\hat{\xi} - \mathcal{T}(x)\right) = A\left(\hat{\xi} - \mathcal{T}(x)\right) \qquad (4.5)$$

which is linear, and it will converge exponentially to 0 when the matrix A is chosen to be Hurwitz.

In summary, existence of a full-state observer with linear error dynamics reduces to two main questions:

i. Solvability of the partial differential equation (4.3)
ii. Injectivity of the mapping $x \mapsto \mathcal{T}(x)$ or $x \mapsto (\mathcal{T}(x), H(x))$.

**i.** has affirmative answer under mild assumptions (see specific results later in this section), whereas for **ii.** to hold, appropriate observability conditions on (4.1) must be imposed (see [10], [11], [12]).

Let's now consider the functional observer, as defined by Definition 1 in Section 2. In the spirit of full-state observer linearization, we seek for a functional observer of the form (2.5) whose dynamics is linear, i.e.

$$\varphi(\xi, y) = A\xi + By \qquad (4.6)$$

Then, condition (2.3) will become the partial differential equation (4.3), but we also need to satisfy condition (2.4), which states that q(x) must be expressible as a function of $\mathcal{T}(x)$ and H(x). This is a functional dependence condition, whose satisfaction depends on the order of the functional observer.

-- If $\nu = n - p$ and the reduced-order full-state observer linearization problem can be solved, this will automatically solve the functional observer linearization problem, *for any* functional q(x).
-- If $\nu < n - p$, the functional dependence requirement from (2.4) may or may not be feasible, depending on q(x) and the observer's dynamics.

In summary, the possibility of extension of exact linearization to functional observers reduces to two main questions:

   **i.** Solvability of the partial differential equation (4.3)
   **ii.** Compatibility of $\mathcal{T}(x)$ with the functional q(x), in the sense of the functional dependence specified through (2.4).

Thus, we see that the solvability problem for the partial differential equation (4.3) is common both for the functional and the full-state observer linearization problems. This problem has been resolved.

For locally Lipschitz F(x) and H(x), local existence of a (weak) solution of (4.3) may be established under very mild assumptions. The following Proposition is an immediate consequence of a Theorem by Andrieu and Praly [10, Theorem 2]:

<u>Proposition 1</u>: *Let $\mathcal{O}$ be an open set. Assume there exists a strictly positive number $\delta$ such that the dynamics $\frac{dx}{dt} = F(x)$ is backward complete within the set $\mathcal{O} + \delta = \{x \in \mathbb{R}^n : \exists \chi \in \mathcal{O} : |x - \chi| < \delta\}$. Then, for every Hurwitz matrix A, there exists a continuous function $\mathcal{T}(x)$, defined on the closure of $\mathcal{O}$, which satisfies (4.3)*.

The above Proposition establishes existence of a solution. It turns out that the solution is unique within the class of locally analytic functions, but under stronger assumptions. The following Proposition is an immediate consequence of Lyapunov's Auxiliary Theorem [18] (see also [11]).

<u>Proposition 1'</u>: *Let $F: \mathbb{R}^n \to \mathbb{R}^n$, $H: \mathbb{R}^n \to \mathbb{R}^p$ be real analytic functions with $F(0) = 0$, $H(0) = 0$ and denote by $\sigma(F)$ the set of eigenvalues of $\frac{\partial F}{\partial x}(0)$. Also, let A and B be $\nu \times \nu$ and $\nu \times p$ matrices respectively. Suppose:*

1. *All the eigenvalues of A are non-resonant with $\sigma(F)$, i.e. no eigenvalue $\lambda_j$ of A is of the form*

$$\lambda_j = \sum_{i=1}^n m_i \kappa_i, \text{ with } \kappa_i \in \sigma(F) \text{ and } m_i \text{ nonnegative integers, not all zero.}$$

2. *0 does not lie in the convex hull of $\sigma(F)$.*

*Then the partial differential equation (4.3) with initial condition $\mathcal{T}(0) = 0$, admits a unique real analytic solution $\mathcal{T}(x)$ in a neighborhood of $x = 0$.*

Because the subproblem of solvability of the partial differential equation (4.3) has been resolved, the focus of the present paper will be on the second subproblem: *under what conditions could the solution $\mathcal{T}(x)$ be compatible with* $q(x)$, *in the sense specified through* (2.4). The goal will be to find conditions to check feasibility of lowering the order of the functional observer, below (n – p).

In the next section we will study a special form of the functional observer linearization problem, where in addition to requiring linear observer dynamics, we will also require linearity of the observer's output map. In particular, we will consider the following:

*Functional Observer Linearization Problem*

*Given a system of the form* (*1.1*), *find a functional observer of the form*

$$\frac{d\hat{\xi}}{dt} = A\hat{\xi} + By$$
$$\hat{z} = C\hat{\xi} + Dy$$
(4.7)

*where A, B, C, D are* $\nu \times \nu$, $\nu \times p$, $1 \times \nu$, $1 \times p$ *matrices respectively, with A having stable eigenvalues. Equivalently, find a continuously differentiable mapping* $\mathcal{T}: \mathbb{R}^n \to \mathbb{R}^\nu$ *such that*

$$\frac{\partial \mathcal{T}}{\partial x} F(x) = A\mathcal{T}(x) + BH(x)$$
(4.3)

*and*

$$q(x) = C\mathcal{T}(x) + DH(x)$$
(4.8)

Assuming that the above problem can be solved, the resulting error dynamics will be linear:

$$\frac{d}{dt}\left(\hat{\xi} - \mathcal{T}(x)\right) = A\left(\hat{\xi} - \mathcal{T}(x)\right)$$
$$\hat{z} - z = C\left(\hat{\xi} - \mathcal{T}(x)\right)$$
(4.9)

from which $\hat{z}(t) - z(t) = Ce^{At}\left(\hat{\xi}(0) - \mathcal{T}(x(0))\right)$. With the matrix A having stable eigenvalues, the effect of the initialization error $\hat{\xi}(0) - \mathcal{T}(x(0))$ will die out, and $\hat{z}(t)$ will approach $z(t)$ asymptotically.

In section 7, we will consider a more general functional observer linearization problem, where the observer dynamics is linear in the observer state but might involve a nonlinear function of y, and the observer output depends on a linear function of the states but might involve a nonlinear function of the observer output and y. In particular, we will consider the following:

*A more general Functional Observer Linearization Problem*

*Given a system of the form (1.1), find a functional observer of the form*

$$\frac{d\hat{\xi}}{dt} = A\hat{\xi} + \mathcal{B}(y) \tag{4.10}$$

$$\hat{z} \text{ solution of } \mathcal{G}(\hat{z}, y) = C\hat{\xi}$$

*where A, C are $\nu \times \nu$, $1 \times \nu$ matrices respectively, with A having stable eigenvalues, and where $\mathcal{B}: \mathbb{R}^p \to \mathbb{R}^\nu$, $\mathcal{G}: \mathbb{R} \times \mathbb{R}^p \to \mathbb{R}$ are continuous functions, with $\mathcal{G}$ invertible with respect to its first argument (i.e., such that the equation $\mathcal{G}(z, y) = \zeta$ is uniquely solvable with respect to z, for every y and $\zeta$).*

*Equivalently, find a continuously differentiable mapping $\mathcal{T}: \mathbb{R}^n \to \mathbb{R}^\nu$ such that*

$$\frac{\partial \mathcal{T}}{\partial x}(x)F(x) = A\mathcal{T}(x) + \mathcal{B}(H(x)) \tag{4.11}$$

*and*

$$\mathcal{G}(q(x), H(x)) = C\mathcal{T}(x) \tag{4.12}$$

Assuming that the foregoing functional observer linearization problem can be solved, the resulting error dynamics will be linear:

$$\frac{d}{dt}\left(\hat{\xi} - \mathcal{T}(x)\right) = A\left(\hat{\xi} - \mathcal{T}(x)\right)$$
$$\mathcal{G}(\hat{z}, H(x)) - \mathcal{G}(z, H(x)) = C\left(\hat{\xi} - \mathcal{T}(x)\right) \tag{4.13}$$

from which $\mathcal{G}\left(\hat{z}(t), H(x(t))\right) - \mathcal{G}\left(z(t), H(x(t))\right) = Ce^{At}\left(\hat{\xi}(0) - \mathcal{T}(x(0))\right)$. With the matrix A having stable eigenvalues, the effect of the initial error $\hat{\xi}(0) - \mathcal{T}(x(0))$ will die out, $\hat{\xi}(t)$ will approach $\mathcal{T}(x(t))$ asymptotically and because of the invertibility assumption on $\mathcal{G}$, $\hat{z}(t)$ will approach $z(t)$

This generalized Functional Observer Linearization Problem will be solved in Section 7.

## 5. NECESSARY AND SUFFICIENT CONDITIONS FOR SOLVABILITY OF THE FUNCTIONAL OBSERVER LINEARIZATION PROBLEM

To be able to develop a practical approach for designing functional observers, it would be helpful to develop criteria to check if for a given set of $\nu$ eigenvalues, there exists a functional observer whose error dynamics is governed by these eigenvalues. This will be done in the present Section for the Functional Observer Linearization Problem

The main result is as follows:

**Proposition 2**: *For a nonlinear system of the form (1.1), there exists a functional observer of the form*

$$\frac{d\hat{\xi}}{dt} = A\hat{\xi} + By$$
$$\hat{z} = C\hat{\xi} + Dy$$
(4.7)

*with the eigenvalues of A being the roots of a given polynomial* $\lambda^\nu + \alpha_1 \lambda^{\nu-1} + \cdots + \alpha_{\nu-1}\lambda + \alpha_\nu$, *if and only if* $L_F^\nu q(x) + \alpha_1 L_F^{\nu-1} q(x) + \cdots + \alpha_{\nu-1} L_F q(x) + \alpha_\nu q(x)$ *is* $\mathbb{R}$-*linear combination of* $H_j(x), L_F H_j(x), \ldots, L_F^\nu H_j(x), j = 1, \cdots, p$, *where* $L_F = \sum_{k=1}^{n} F_k(x) \frac{\partial}{\partial x_k}$ *denotes the Lie derivative operator.*

**Proof**: i) *Necessity*

Suppose that there exists a functional observer of the form (4.7) for the system (1.1). Then, condition (4.3) will be satisfied for some $\mathcal{T}(x) = \begin{bmatrix} \mathcal{T}_1(x) \\ \vdots \\ \mathcal{T}_\nu(x) \end{bmatrix}$. Using the Lie derivative operator notation, this condition may be written component-wise as $\begin{bmatrix} L_F \mathcal{T}_1(x) \\ \vdots \\ L_F \mathcal{T}_\nu(x) \end{bmatrix} = A \begin{bmatrix} \mathcal{T}_1(x) \\ \vdots \\ \mathcal{T}_\nu(x) \end{bmatrix} + \begin{bmatrix} B_1 H(x) \\ \vdots \\ B_\nu H(x) \end{bmatrix}$.

where $B_1, \cdots, B_\nu$ denote the rows of matrix B. Applying the Lie derivative operator $L_F$ to each component in the above equation (k – 1) times, we find that for k = 2, 3, …

$$\begin{bmatrix} L_F^k \mathcal{T}_1(x) \\ \vdots \\ L_F^k \mathcal{T}_\nu(x) \end{bmatrix} = A^k \begin{bmatrix} \mathcal{T}_1(x) \\ \vdots \\ \mathcal{T}_\nu(x) \end{bmatrix} + A^{k-1} \begin{bmatrix} B_1 H(x) \\ \vdots \\ B_\nu H(x) \end{bmatrix} + A^{k-2} \begin{bmatrix} L_F(B_1 H(x)) \\ \vdots \\ L_F(B_\nu H(x)) \end{bmatrix} + \cdots + \begin{bmatrix} L_F^{k-1}(B_1 H(x)) \\ \vdots \\ L_F^{k-1}(B_\nu H(x)) \end{bmatrix}$$

from which we can calculate:

$$\begin{bmatrix} (L_F^\nu + \alpha_1 L_F^{\nu-1} + \cdots + \alpha_{\nu-1} L_F + \alpha_\nu I) \mathcal{T}_1(x) \\ \vdots \\ (L_F^\nu + \alpha_1 L_F^{\nu-1} + \cdots + \alpha_{\nu-1} L_F + \alpha_\nu I) \mathcal{T}_\nu(x) \end{bmatrix} = (A^{\nu-1} + \alpha_1 A^{\nu-2} + \cdots + \alpha_{\nu-1} I) \begin{bmatrix} B_1 H(x) \\ \vdots \\ B_\nu H(x) \end{bmatrix}$$

$$+ (A^{\nu-2} + \alpha_1 A^{\nu-3} + \cdots + \alpha_{\nu-2} I) \begin{bmatrix} L_F(B_1 H(x)) \\ \vdots \\ L_F(B_\nu H(x)) \end{bmatrix} + \cdots + (A + \alpha_1 I) \begin{bmatrix} L_F^{\nu-2}(B_1 H(x)) \\ \vdots \\ L_F^{\nu-2}(B_\nu H(x)) \end{bmatrix} + \begin{bmatrix} L_F^{\nu-1}(B_1 H(x)) \\ \vdots \\ L_F^{\nu-1}(B_\nu H(x)) \end{bmatrix}$$

Applying the operator $\left(L_F^v + \alpha_1 L_F^{v-1} + \cdots + \alpha_{v-1} L_F + \alpha_v I\right)$ on (4.8) and using the previous expression, we conclude that

$$\left(L_F^v + \alpha_1 L_F^{v-1} + \cdots + \alpha_{v-1} L_F + \alpha_v I\right) q(x) = \left(CA^{v-1} + \alpha_1 CA^{v-2} + \cdots + \alpha_{v-1} C\right) \begin{bmatrix} B_1 H(x) \\ \vdots \\ B_v H(x) \end{bmatrix}$$

$$+ \left(CA^{v-2} + \cdots + \alpha_{v-2} C\right) \begin{bmatrix} L_F(B_1 H(x)) \\ \vdots \\ L_F(B_v H(x)) \end{bmatrix} + \cdots + \left(CA + \alpha_1 C\right) \begin{bmatrix} L_F^{v-2}(B_1 H(x)) \\ \vdots \\ L_F^{v-2}(B_v H(x)) \end{bmatrix} + C \begin{bmatrix} L_F^{v-1}(B_1 H(x)) \\ \vdots \\ L_F^{v-1}(B_v H(x)) \end{bmatrix}$$

$$+ L_F^v(DH(x)) + \alpha_1 L_F^{v-1}(DH(x)) + \cdots + \alpha_{v-1} L_F(DH(x)) + \alpha_v DH(x)$$

or

$$L_F^v q(x) + \alpha_1 L_F^{v-1} q(x) + \cdots + \alpha_{v-1} L_F q(x) + \alpha_v q(x)$$
$$= L_F^v(\beta_0 H(x)) + L_F^{v-1}(\beta_1 H(x)) + \cdots + L_F(\beta_{v-1} H(x)) + \beta_v H(x) \quad (5.1)$$
$$= \sum_{j=1}^{p} \left(\beta_{0_j} L_F^v H_j(x) + \beta_{1_j} L_F^{v-1} H_j(x) + \cdots + \beta_{(v-1)_j} L_F H_j(x) + \beta_{v_j} H_j(x)\right)$$

where

$$\begin{aligned} \beta_0 &= D \\ \beta_1 &= CB + \alpha_1 D \\ \beta_2 &= CAB + \alpha_1 CB + \alpha_2 D \\ &\vdots \\ \beta_{v-1} &= CA^{v-2} B + \cdots + \alpha_{v-2} CB + \alpha_{v-1} D \\ \beta_v &= CA^{v-1} B + \alpha_1 CA^{v-2} B + \cdots + \alpha_{v-1} CB + \alpha_v D \end{aligned} \quad (5.2)$$

This proves that $L_F^v q(x) + \alpha_1 L_F^{v-1} q(x) + \cdots + \alpha_{v-1} L_F q(x) + \alpha_v q(x)$ is $\mathbb{R}$-linear combination of $H_j(x), L_F H_j(x), \ldots, L_F^v H_j(x), j = 1, \cdots, p$.

ii) *Sufficiency*: Suppose that $L_F^v q(x) + \alpha_1 L_F^{v-1} q(x) + \cdots + \alpha_{v-1} L_F q(x) + \alpha_v q(x)$ is $\mathbb{R}$-linear combination of $H_j(x), L_F H_j(x), \ldots, L_F^v H_j(x), j = 1, \cdots, p$, i.e. there exist constant row vectors $\beta_0, \beta_1, \cdots, \beta_v \in \mathbb{R}^p$ such that (5.1) holds.

Consider the partial differential equation:

$$\frac{\partial \mathcal{T}}{\partial x}(x)F(x) = \begin{bmatrix} 0 & 0 & \cdots & 0 & -\alpha_\nu \\ 1 & 0 & \cdots & 0 & -\alpha_{\nu-1} \\ 0 & 1 & \cdots & 0 & -\alpha_{\nu-2} \\ \vdots & \vdots & & \vdots & \vdots \\ 0 & 0 & \cdots & 1 & -\alpha_1 \end{bmatrix} \mathcal{T}(x) + \begin{bmatrix} \beta_\nu - \alpha_\nu \beta_0 \\ \beta_{\nu-1} - \alpha_{\nu-1}\beta_0 \\ \beta_{\nu-2} - \alpha_{\nu-2}\beta_0 \\ \vdots \\ \beta_1 - \alpha_1 \beta_0 \end{bmatrix} H(x) \quad (5.3)$$

It is straightforward to verify that

$$\mathcal{T}(x) = \begin{bmatrix} L_F^{\nu-1} q(x) + \alpha_1 L_F^{\nu-2} q(x) + \cdots + \alpha_{\nu-1} q(x) - L_F^{\nu-1}(\beta_0 H(x)) - L_F^{\nu-2}(\beta_1 H(x)) - \cdots - \beta_{\nu-1} H(x) \\ \vdots \\ L_F q(x) + \alpha_1 q(x) - L_F(\beta_0 H(x)) - \beta_1 H(x) \\ q(x) - \beta_0 H(x) \end{bmatrix}$$

(5.4)

satisfies the PDE (5.3) and we see that its $\nu$-th component is $\mathcal{T}_\nu(x) = q(x) - \beta_0 H(x)$, therefore,

$$q(x) = \begin{bmatrix} 0 & 0 & \cdots & 0 & 1 \end{bmatrix} \mathcal{T}(x) + \beta_0 H(x) \quad (5.5)$$

Hence $\mathcal{T}(x)$ given by (5.4) satisfies conditions (4.3) and (4.8) for the solution of the Functional Observer Linearization Problem, and system (4.7) with

$$A = \begin{bmatrix} 0 & 0 & \cdots & 0 & -\alpha_\nu \\ 1 & 0 & \cdots & 0 & -\alpha_{\nu-1} \\ 0 & 1 & \cdots & 0 & -\alpha_{\nu-2} \\ \vdots & \vdots & & \vdots & \vdots \\ 0 & 0 & \cdots & 1 & -\alpha_1 \end{bmatrix}, \quad B = \begin{bmatrix} \beta_\nu - \alpha_\nu \beta_0 \\ \beta_{\nu-1} - \alpha_{\nu-1}\beta_0 \\ \beta_{\nu-2} - \alpha_{\nu-2}\beta_0 \\ \vdots \\ \beta_1 - \alpha_1 \beta_0 \end{bmatrix}$$

(5.6)

$$C = \begin{bmatrix} 0 & 0 & \cdots & 0 & 1 \end{bmatrix}, \quad D = \beta_0$$

is a functional observer. This completes the proof. $\square$

It is important to emphasize that the sufficiency part of the proof is constructive, and it immediately leads to a design method for the functional observer:

*Once a set of constant row vectors* $\beta_0, \beta_1, \cdots, \beta_\nu \in \mathbb{R}^p$ *have been found to satisfy (5.1) for a specific characteristic polynomial* $\lambda^\nu + \alpha_1 \lambda^{\nu-1} + \cdots + \alpha_{\nu-1}\lambda + \alpha_\nu$, *formula (5.6) immediately gives the* A, B, C *and* D *matrices of the linear functional observer.*

Also, it should be noted that there may be multiple sets of $\beta_0, \beta_1, \cdots, \beta_\nu \in \mathbb{R}^p$ that satisfy (5.1), leading to multiple solutions for the functional observer linearization problem.

*Chemical reactor with hazardous reactants*

Consider a non-isothermal Continuous Stirred Tank Reactor (CSTR) where an exothermic chemical reaction A+B → C+D takes place. The reactor is cooled through a cooling jacket. The reactor dynamics can be modelled through standard component mas balances and energy balances, assuming constant volume and constant thermophysical properties, as follows ([19, 20]):

$$\begin{aligned}
\frac{dc_A}{dt} &= \frac{F}{V}(c_{A_{in}} - c_A) - \mathcal{R}(c_A, c_B, \theta) \\
\frac{dc_B}{dt} &= \frac{F}{V}(c_{B_{in}} - c_B) - \mathcal{R}(c_A, c_B, \theta) \\
\frac{d\theta}{dt} &= \frac{F}{V}(\theta_{in} - \theta) + \frac{(-\Delta H)_R}{\rho c_p} \mathcal{R}(c_A, c_B, \theta) - \frac{UA}{\rho c_p V}(\theta - \theta_J) \\
\frac{d\theta_J}{dt} &= \frac{F_J}{V_J}(\theta_{J_{in}} - \theta_J) + \frac{UA}{\rho_J c_{p_J} V_J}(\theta - \theta_J)
\end{aligned} \tag{5.7}$$

where $c_A$ and $c_B$ are the concentrations of species A and B respectively in the reacting mixture, $\theta$ and $\theta_J$ are the temperatures of the reacting mixture and the jacket fluid respectively; these are the system states. The function $\mathcal{R}(c_A, c_B, \theta)$ represents the reaction rate and it is a given algebraic function, specified in terms of an empirical correlation. The rest of the symbols represent constant parameters: $c_{A_{in}}$ and $c_{B_{in}}$ are the feed concentrations of species A and B respectively, $F$ and $F_J$ are the feed and coolant flowrates respectively, $V$ and $V_J$ are the reactor volume and cooling jacket volume respectively, $(-\Delta H)_R$ is the heat of reaction, $\rho, c_p$ and $\rho_J, c_{p_J}$ are the densities and heat capacities of the reactor contents and cooling fluid respectively, $U$ and $A$ are the overall heat transfer coefficient and heat transfer area respectively.

When the reactants A and B are potentially hazardous, special precautions are taken in terms of using relatively dilute feeds and the reaction taking place at a relatively low temperature. In terms of monitoring the operation of the reactor, the temperature $\theta$ of the reacting mixture as well as the total sum of hazardous chemicals' concentrations $c_A + c_B$ are critical quantities to be monitored. Temperature is easy and inexpensive to measure, but concentrations generally need to be estimated from temperature measurements. Consider therefore the problem of building an observer for the dynamic system (5.7), driven by the temperature measurements

$$\begin{aligned} y_1 &= \theta \\ y_2 &= \theta_J \end{aligned} \tag{5.8}$$

the objective being to estimate the sum of the reactant concentrations

$$z = c_A + c_B \tag{5.9}$$

Liquid-phase oxidation reactions are a very important class of chemical reactions that are notorious for being highly exothermic and for involving serious safety threats. One well-studied example is the reaction of N-methyl pyridine (A) with hydrogen peroxide (B) in the presence of a catalyst [19].

For this reaction, the reaction rate expression is (see [19]):

$$\mathcal{R}(c_A, c_B, \theta) = \frac{A_1 e^{-\frac{E_1}{\theta}} A_2 e^{-\frac{E_2}{\theta}} c_A c_B Z}{1 + A_2 e^{-\frac{E_2}{\theta}} c_B} + A_3 e^{-\frac{E_3}{\theta}} c_A c_B \qquad (5.10)$$

where $A_1, A_2, A_3$ and $E_1, E_2, E_3$ are the reaction rate parameters, pre-exponential factors and rescaled activation energies respectively, and $Z$ is the catalyst concentration (constant).
To derive a functional observer, it is convenient to perform appropriate translation of axes to shift the equilibrium point to the origin. In particular, defining $c'_A = c_A - c_{A,s}$, $c'_B = c_B - c_{B,s}$, $\theta' = \theta - \theta_s$, $\theta'_J = \theta_J - \theta_{J,s}$, where $(c_{A,s}, c_{B,s}, \theta_s, \theta_{J,s})$ is the steady state (equilibrium point) of the reactor. For the above system, a scalar functional observer can then be built ($v=1$), with the necessary and sufficient conditions (5.1) being satisfied for:

$$\beta_0 = \left[-\frac{2\rho c_p}{(-\Delta H)_R} \quad 0\right], \quad \beta_1 = \left[-\frac{2\rho c_p}{(-\Delta H)_R}\left(\frac{F}{V} + \frac{UA}{\rho c_p V}\right) \quad \frac{2UA}{(-\Delta H)_R V}\right], \quad \alpha_1 = \frac{F}{V} \qquad (5.11)$$

The corresponding transformation map is $\mathcal{T}(c'_A, c'_B, \theta', \theta'_J) = c'_A + c'_B + \frac{2\rho c_p}{(-\Delta H)_R}\theta'$, and the resulting functional observer is given by:

$$\frac{d\hat{\xi}}{dt} = -\frac{F}{V}\hat{\xi} - \frac{2UA}{(-\Delta H)_R V}(y'_1 - y'_2)$$

$$\hat{z} = \hat{\xi} - \frac{2\rho c_p}{(-\Delta H)_R} y'_1 \qquad (5.12)$$

For the following parameter values (see [19]):

$c_{A,in} = 4\frac{mol}{l}$, $c_{B,in} = 3\frac{mol}{l}$, $\theta_{in} = 333\, K$, $\theta_{J,in} = 300\, K$, $F = 0.02\frac{l}{min}$, $F_J = 1\frac{l}{min}$,
$V = 1\, l$, $V_J = 3 \times 10^{-2}\, l$, $A_1 = e^{8.08}\, l\, mol^{-1} s^{-1}$, $A_2 = e^{28.12}\, l\, mol^{-1} s^{-1}$,
$A_3 = e^{25.12}\, l\, mol^{-1}$, $E_1 = 3952\, K$, $E_2 = 7927\, K$, $E_3 = 12989\, K$, $\Delta H_R = -160\frac{kJ}{mol}$,
$\rho = 1200\frac{g}{l}$, $\rho_J = 1200\frac{g}{l}$, $c_{p_J} = 3.4\frac{J}{gK}$, $c_p = 3.4\frac{J}{gK}$, $UA = 0.942\frac{W}{K}$, $Z = 0.0021\frac{mol}{l}$

the corresponding reactor steady state is:

$c_{A,s} = 1.211\frac{mol}{l}$, $c_{B,s} = 0.211\frac{mol}{l}$, $\theta_s = 386.20\, K$, $\theta_{J,s} = 300.02\, K$,

and we have simulated the reactor start-up, under the following initial conditions:
$c_A(0) = 0$, $c_B(0) = 0$, $\theta(0) = 300\, K$, $\theta_J(0) = 300\, K$.

Figure 1 compares the functional observer's estimate $\hat{c}_A + \hat{c}_B = \hat{z} + c_{A,s} + c_{B,s}$ to the system's total reactant concentration $c_A + c_B$, and provides a plot of the corresponding estimation error, when the initialization error is $1\frac{mol}{l}$.

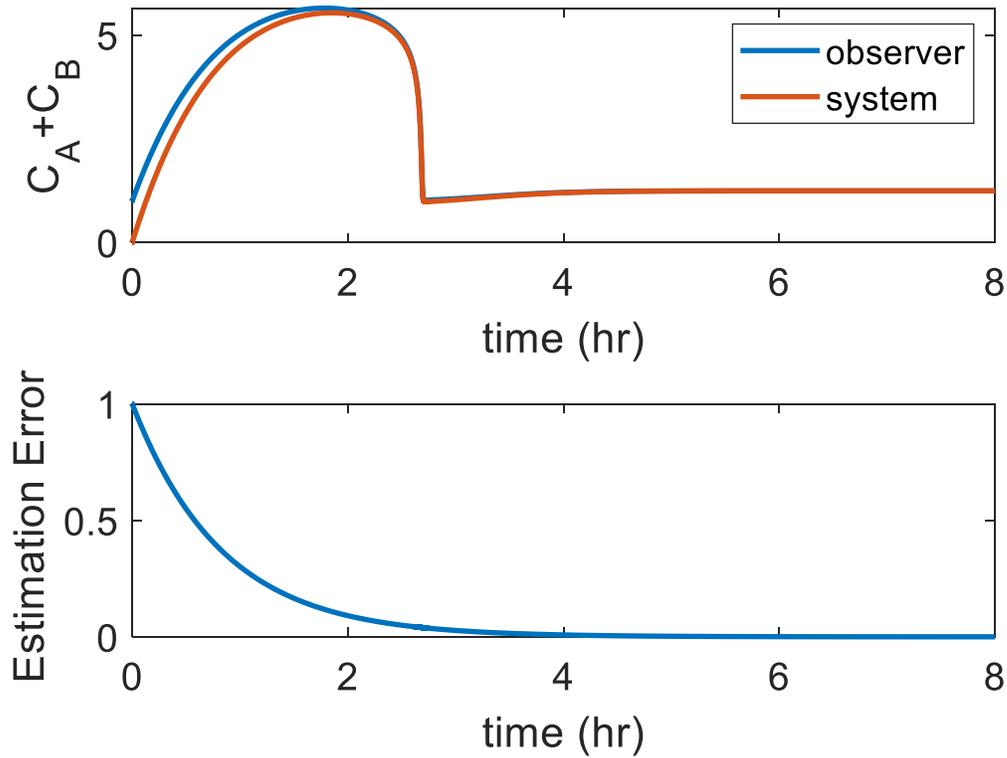

Figure 1: (a) System's and observer's response for $\hat{\xi}(0) - \mathcal{T}(c'_A(0), c'_B(0), \theta'(0), \theta'_J(0)) = 1\frac{mol}{\ell}$

(b) Estimation error $\hat{z}(t) - z(t)$

## 6. LOWER-ORDER FUNCTIONAL OBSERVERS FOR LINEAR SYSTEMS

The results of the previous section can now be specialized to linear time-invariant systems. The following is a Corollary to Proposition 2.

Proposition 3: *For a linear time-invariant system of the form*

$$\frac{dx}{dt} = Fx$$
$$y = Hx \qquad (2.7)$$
$$z = qx$$

there exists a functional observer of the form

$$\frac{d\hat{\xi}}{dt} = A\hat{\xi} + By \qquad (2.8)$$
$$\hat{z} = C\hat{\xi} + Dy$$

*with the eigenvalues of* A *being the roots of a given polynomial* $\lambda^\nu + \alpha_1\lambda^{\nu-1} + \cdots + \alpha_{\nu-1}\lambda + \alpha_\nu$,

*if and only if*

$$(qF^\nu + \alpha_1 qF^{\nu-1} + \cdots + \alpha_{\nu-1} qF + \alpha_\nu q) \in \text{span}\{H_j, H_j F, \ldots, H_j F^\nu, \ j = 1, \cdots, p\} \quad (6.1)$$

The above Proposition provides a simple and easy-to-check feasibility criterion for a lower-order functional observer with a pre-specified set of eigenvalues governing the error dynamics. Moreover, an immediate consequence of the Proposition 3 is the following:

<u>Corollary</u>: *Consider a linear time-invariant system of the form (2.7) with observability index $\nu_o$. Then, there exists a functional observer of the form (2.8) of order $\nu = \nu_o - 1$ and arbitrarily assigned eigenvalues.*

The result of the Corollary is exactly Luenberger's result for functional observers for linear time-invariant systems [1, 2], which was derived through a different approach.

## 7. SOLVABILITY OF A MORE GENERAL FUNCTIONAL OBSERVER LINEARIZATION PROBLEM

Consider now the more general functional observer linearization problem defined in Section 4. An observer of the form (4.10) is sought.

Using the same method as in the proof of Proposition 2, we can prove the following:

<u>Proposition 2′</u>: *For a system of the form (1.1), there exists a functional observer of the form (4.10) with the eigenvalues of A being the roots of a given polynomial $\lambda^\nu + \alpha_1 \lambda^{\nu-1} + \cdots + \alpha_{\nu-1} \lambda + \alpha_\nu$, if and only if there exist functions $\mathcal{B}_0 : \mathbb{R} \times \mathbb{R}^p \to \mathbb{R}$, invertible with respect to its first argument, and $\mathcal{B}_1, \cdots, \mathcal{B}_\nu : \mathbb{R}^p \to \mathbb{R}$ such that*

$$\left(L_F^\nu + \alpha_1 L_F^{\nu-1} + \cdots + \alpha_{\nu-1} L_F + \alpha_\nu I\right) \mathcal{B}_0(q(x), H(x)) = \mathcal{B}_1(H(x)) + L_F(\mathcal{B}_2(H(x))) + \cdots + L_F^{\nu-1}(\mathcal{B}_\nu(H(x))) \quad (7.1)$$

*Proof*: i) *Necessity*: Following exactly the same steps as in the proof of Proposition 2, we can first conclude from (4.11) that

$$\begin{bmatrix} \left(L_F^\nu + \alpha_1 L_F^{\nu-1} + \cdots + \alpha_{\nu-1} L_F + \alpha_\nu I\right) \mathcal{T}_1(x) \\ \vdots \\ \left(L_F^\nu + \alpha_1 L_F^{\nu-1} + \cdots + \alpha_{\nu-1} L_F + \alpha_\nu I\right) \mathcal{T}_\nu(x) \end{bmatrix} = \left(A^{\nu-1} + \alpha_1 A^{\nu-2} + \cdots + \alpha_{\nu-1} I\right) \begin{bmatrix} \mathcal{B}_1(H(x)) \\ \vdots \\ \mathcal{B}_\nu(H(x)) \end{bmatrix}$$

$$+ \left(A^{\nu-2} + \alpha_1 A^{\nu-3} + \cdots + \alpha_{\nu-2} I\right) \begin{bmatrix} L_F \mathcal{B}_1(H(x)) \\ \vdots \\ L_F \mathcal{B}_\nu(H(x)) \end{bmatrix} + \cdots + (A + \alpha_1 I) \begin{bmatrix} L_F^{\nu-2} \mathcal{B}_1(H(x)) \\ \vdots \\ L_F^{\nu-2} \mathcal{B}_\nu(H(x)) \end{bmatrix} + \begin{bmatrix} L_F^{\nu-1} \mathcal{B}_1(H(x)) \\ \vdots \\ L_F^{\nu-1} \mathcal{B}_\nu(H(x)) \end{bmatrix}$$

Then, applying the operator $\left(L_F^\nu + \alpha_1 L_F^{\nu-1} + \cdots + \alpha_{\nu-1} L_F + \alpha_\nu I\right)$ on (4.12) and using the previous expression, we conclude that

$$\left(L_F^v + \alpha_1 L_F^{v-1} + \cdots + \alpha_{v-1} L_F + \alpha_v I\right) \mathcal{G}(q(x), H(x)) = \left(CA^{v-1} + \alpha_1 CA^{v-2} + \cdots + \alpha_{v-1} C\right) \begin{bmatrix} \mathcal{B}_1(H(x)) \\ \vdots \\ \mathcal{B}_v(H(x)) \end{bmatrix}$$

$$+ \left(CA^{v-2} + \cdots + \alpha_{v-2} C\right) \begin{bmatrix} L_F \mathcal{B}_1(H(x)) \\ \vdots \\ L_F \mathcal{B}_v(H(x)) \end{bmatrix} + \cdots + \left(CA + \alpha_1 C\right) \begin{bmatrix} L_F^{v-2} \mathcal{B}_1(H(x)) \\ \vdots \\ L_F^{v-2} \mathcal{B}_v(H(x)) \end{bmatrix} + C \begin{bmatrix} L_F^{v-1} \mathcal{B}_1(H(x)) \\ \vdots \\ L_F^{v-1} \mathcal{B}_v(H(x)) \end{bmatrix}$$

or

$$\left(L_F^v + \alpha_1 L_F^{v-1} + \cdots + \alpha_{v-1} L_F + \alpha_v I\right) \mathcal{Z}_0(q(x), H(x)) = \mathcal{Z}_1(H(x)) + L_F\left(\mathcal{Z}_2(H(x))\right) + \cdots + L_F^{v-1}\left(\mathcal{Z}_v(H(x))\right)$$

where

$$\begin{aligned}
\mathcal{Z}_0(z, y) &= \mathcal{G}(z, y) \\
\mathcal{Z}_1(y) &= \left(CA^{v-1} + \alpha_1 CA^{v-2} + \cdots + \alpha_{v-1} C\right) \mathcal{B}(y) \\
\mathcal{Z}_2(y) &= \left(CA^{v-2} + \cdots + \alpha_{v-2} C\right) \mathcal{B}(y) \\
&\vdots \\
\mathcal{Z}_{v-1}(y) &= \left(CA + \alpha_1 C\right) \mathcal{B}(y) \\
\mathcal{Z}_v(y) &= C \mathcal{B}(y)
\end{aligned} \qquad (7.2)$$

ii) *Sufficiency*: Assuming that (7.1) holds, we can follow the same steps as in the proof of Proposition 2 and prove that

$$\mathcal{T}(x) = \begin{bmatrix} \left(L_F^{v-1} + \alpha_1 L_F^{v-2} + \cdots + \alpha_{v-2} L_F + \alpha_{v-1} I\right) \mathcal{Z}_0(q(x), H(x)) - \mathcal{Z}_2(H(x)) - \cdots - L_F^{v-2}\left(\mathcal{Z}_v(H(x))\right) \\ \vdots \\ \left(L_F^2 + \alpha_1 L_F + \alpha_2 I\right) \mathcal{Z}_0(q(x), H(x)) - \mathcal{Z}_{v-1}(H(x)) - L_F\left(\mathcal{Z}_v(H(x))\right) \\ \left(L_F + \alpha_1 I\right) \mathcal{Z}_0(q(x), H(x)) - \mathcal{Z}_v(H(x)) \\ \mathcal{Z}_0(q(x), H(x)) \end{bmatrix}$$

(7.3)

satisfies both conditions (4.11) and (4.12) of the generalized functional observer linearization problem, with

$$A = \begin{bmatrix} 0 & 0 & \cdots & 0 & -\alpha_\nu \\ 1 & 0 & \cdots & 0 & -\alpha_{\nu-1} \\ 0 & 1 & \cdots & 0 & -\alpha_{\nu-2} \\ \vdots & \vdots & & \vdots & \vdots \\ 0 & 0 & \cdots & 1 & -\alpha_1 \end{bmatrix}, \quad \mathcal{B}(y) = \begin{bmatrix} \mathcal{B}_1(y) \\ \mathcal{B}_2(y) \\ \mathcal{B}_3(y) \\ \vdots \\ \mathcal{B}_\nu(y) \end{bmatrix}$$

(7.4)

$$C = \begin{bmatrix} 0 & 0 & \cdots & 0 & 1 \end{bmatrix}, \quad \mathcal{G}(z, y) = \mathcal{B}_0(z, y)$$

Therefore, system (7.1) with A, $\mathcal{B}(\cdot)$, C, $\mathcal{G}(\cdot,\cdot)$ given by (8.4) is a functional observer. □

*Example*: Consider the following system

$$\frac{dx_1}{dt} = -x_1 - x_3^6$$

$$\frac{dx_2}{dt} = \sin(x_1^2) - x_3^2 - x_2$$

$$\frac{dx_3}{dt} = -x_3 + x_1 x_2 - \frac{1}{1 + x_3^2} \quad (7.5)$$

$$y = x_3^2$$

$$z = x_1 + x_3^4$$

A scalar ($\nu=1$) linear functional observer can be built with

$$\mathcal{B}_0(z, y) = z - y^2$$
$$\mathcal{B}_1(y) = -y^3 \quad (7.6)$$

which satisfy condition (7.1) with and $\alpha_1 = 1$. The resulting functional observer (from (4.10) and (7.4)) is

$$\frac{d\hat{\xi}}{dx} = -\hat{\xi} - y^3$$
$$\hat{z} = \hat{\xi} + y^2 \quad (7.7)$$

## 8. CONCLUSION

The present work has developed a direct generalization of Luenberger's functional observers to nonlinear systems. It has formulated notions of exact linearization for functional observer design and has derived specific criteria for linearization to be feasible, including a simple formula for the resulting functional observer. Unlike full-order and reduced-order state observers that can be designed to have linearizable error dynamics for almost all real analytic nonlinear systems in the Poincaré domain without restrictions, functional observers can only be linearized under rather restrictive conditions.

ACKNOWLEDGMENT

Financial support from the National Science Foundation through the grant CBET-1706201 is gratefully acknowledged.

REFERENCES

1. Luenberger, D., *Observers for multivariable systems.* IEEE Transactions on Automatic Control, 1966. **11**(2): p. 190-197.
2. Luenberger, D., *An introduction to observers.* IEEE Transactions on automatic control, 1971. **16**(6): p. 596-602.
3. Chen, C., *Linear system theory and design: Oxford University Press.* New York, USA, 1999.
4. Darouach, M., *Existence and design of functional observers for linear systems.* IEEE Transactions on Automatic Control, 2000. **45**(5): p. 940-943.
5. Fernando, T.L., H.M. Trinh, and L. Jennings, *Functional observability and the design of minimum order linear functional observers.* IEEE Transactions on Automatic Control, 2010. **55**(5): p. 1268-1273.
6. Korovin, S., I. Medvedev, and V. Fomichev. *Minimum dimension of a functional observer with a given convergence rate*. in *Doklady Mathematics*. 2008. Springer.
7. Korovin, S., I. Medvedev, and V. Fomichev, *Minimum-order functional observers.* Computational Mathematics and Modeling, 2010. **21**(3): p. 275-296.
8. Trinh, H., T. Fernando, and S. Nahavandi, *Partial-state observers for nonlinear systems.* IEEE transactions on automatic control, 2006. **51**(11): p. 1808-1812.
9. Tsui, C.-C., *What is the minimum function observer order?* Journal of the Franklin Institute, 1998. **335**(4): p. 623-628.
10. Andrieu, V. and L. Praly, *On the Existence of a Kazantzis--Kravaris/Luenberger Observer.* SIAM Journal on Control and Optimization, 2006. **45**(2): p. 432-456.
11. Kazantzis, N. and C. Kravaris, *Nonlinear observer design using Lyapunov's auxiliary theorem.* Systems & Control Letters, 1998. **34**(5): p. 241-247.
12. Kazantzis, N., C. Kravaris, and R.A. Wright, *Nonlinear observer design for process monitoring.* Industrial & engineering chemistry research, 2000. **39**(2): p. 408-419.
13. Kreisselmeier, G. and R. Engel, *Nonlinear observers for autonomous Lipschitz continuous systems.* IEEE Transactions on Automatic Control, 2003. **48**(3): p. 451-464.
14. Krener, A.J. and A. Isidori, *Linearization by output injection and nonlinear observers.* Systems & Control Letters, 1983. **3**(1): p. 47-52.
15. Krener, A.J. and W. Respondek, *Nonlinear observers with linearizable error dynamics.* SIAM Journal on Control and Optimization, 1985. **23**(2): p. 197-216.
16. Krener, A.J. and M. Xiao, *Nonlinear observer design in the Siegel domain.* SIAM Journal on Control and Optimization, 2002. **41**(3): p. 932-953.
17. Krener, A.J. and M. Xiao, *Nonlinear observer design for smooth systems.* Chaos in automatic control, 2005: p. 411-422.
18. Lyapunov, A.M., *The general problem of the stability of motion.* International journal of control, 1992. **55**(3): p. 531-534.

19. Cui, X., M.S. Mannan, and B.A. Wilhite, *Towards efficient and inherently safer continuous reactor alternatives to batch-wise processing of fine chemicals: CSTR nonlinear dynamics analysis of alkylpyridines N-oxidation.* Chemical Engineering Science, 2015. **137**: p. 487-503.
20. Fogler, H.S., *Elements of chemical reaction engineering*. 4th ed. Prentice Hall PTR international series in the physical and chemical engineering sciences. 2006, Upper Saddle River, NJ: Prentice Hall PTR.